\documentstyle[prl,multicol,aps,epsfig]{revtex}

\newcommand{\br}{{\bf r}}
\newcommand{\bv}{{\bf v}}
\newcommand{\cf}{{\cal F}}
\newcommand{\cg}{{\cal G}}
\newcommand{\Pe}{{\rm Pe}}
\newcommand{\Da}{{\rm Da}}
\newcommand{\BE}{\begin{equation}}
\newcommand{\EE}{\end{equation}}
\newcommand{\BA}{\begin{eqnarray}}
\newcommand{\EA}{\end{eqnarray}}

\begin{document}
\draft

\title{Excitable media in open and closed chaotic flows}
\author{Zolt\'an Neufeld$^1$, Crist\'obal L\'{o}pez$^2$,
Emilio Hern\'{a}ndez-Garc\'{\i}a$^3$ and  Oreste Piro$^3$}

\address{$^1$Department of Applied Mathematics and Theoretical Physics,
University of Cambridge, Silver Street, Cambridge CB3 9EW, UK \\
$^2$Dipartimento di Fisica, Universit\`{a} di Roma `La Sapienza',
P.le A. Moro 2, I-00185, Roma, Italy \\ $^{3}$ IMEDEA (CSIC-UIB)
Instituto Mediterr\'aneo de Estudios Avanzados, Campus Universitat
de les Illes Balears, E-07071 Palma de Mallorca, Spain}
\date{\today}
\maketitle

\begin{abstract}
We investigate the response of an excitable medium to a localized
perturbation in the presence of a two-dimensional smooth chaotic
flow. Two distinct types of flows are numerically considered: {\it
open} and {\it closed}. For both of them three distinct regimes
are found, depending on the relative strengths of the stirring and
the rate of the excitable reaction. In order to clarify and
understand the role of the many competing mechanisms present,
simplified models of the process are introduced. They are
one-dimensional baker-map models for the flow and a
one-dimensional approximation for the transverse profile of the
filaments.
\end{abstract}


\begin{multicols}{2}

\section{Introduction}
\label{sec:intro}
Excitable media \cite{excitable,Murray,excitable1} are extended
systems exhibiting a variety of pattern formation phenomena. They
are often of chemical or biological nature, although they can also
be found in other contexts\cite{BK,laser}. Each spatial point in
an excitable medium is described by a dynamical system in which
{\sl activator} and {\sl inhibitor} variables can be identified.
The {\sl activator} displays some kind of autocatalytic growth
behavior, but the presence of the {\sl inhibitor} controls it so
that the dynamical system has a stable fixed point as unique
global attractor. The essence of the excitability phenomenon is
the presence of a threshold, such that if the system is perturbed
above it, the system variables reach the stable fixed point only
after a large excursion in phase space. This behavior usually
appears when the activator has a temporal response much faster
than the inhibitor, which then takes some time before stopping the
growth of the activator.

When different parts of a system are coupled diffusively, a local
perturbation excites neighboring points, and as a result, the
excitation propagates through the system as a wave (called {\sl
autowave} or {\sl front}). This is a global phenomenon in the
sense that all the points in the system will be reached by the
wave and thus experience the excitation-deexcitation cycle, but is
non-coherent, since only a small part of the system (the front) is
excited at each time.

In many situations the excitable dynamics takes place in a fluid
environment. One such example is the Belousov-Zhabotinsky (BZ)
reaction \cite{BZ,kapral}, intensively investigated in laboratory
experiments. Another example is the population competition
occurring in oceans or lakes between different plankton species:
Truscott and Brindley \cite{plankton} identified phytoplankton as
the fast activator and zooplankton as the slow inhibitor in models
of biological aquatic population dynamics. In such situations,
different parts of the system interact not only via diffusion, but
advective transport is present, and it can play also an important
role\cite{epstein}. One of the most obvious effects of stirring by
the flow is that the concentrations would become more mixed and,
for fast enough stirring, the whole extended system will behave
homogeneously. Another effect, in this case present for slow
stirring, is that the fronts would be deformed by the flow and
eventually be broken \cite{shear}. A study covering the full range
of stirring intensities was performed in \cite{PRL}, in the
framework of flows leading to chaotic advection\cite{Aref2002}.
Among other results, it was shown that there is an intermediate
range of stirring for which the whole system excites coherently.

In this Paper we further analyze the situation addressed in
\cite{PRL}, that is, excitable media under the effect of chaotic
advection, and extend it by considering also stirring by open
flows. As a striking result, we find situations in which
excitation persists indefinitely in the system when stirred by an
open flow, whereas the excitation process is a transient both
under closed flows and in the absence of stirring. Additionally, a
number of simplified one-dimensional models are introduced and
used to gain insight and analytical predictions on the dynamical
processes involved. We mention that studies in the same spirit
than ours but for the different case of chemical reactions of
autocatalytic or bistable type can be found in \cite{ZoltanFocus}.

The Paper is organized as follows: In Section \ref{sec:rad} we
present the basic framework and the chemical and two-dimensional
flow models (closed and open) to be used. Section
\ref{sec:numerics} describes numerical results for them. Section
\ref{sec:reduced} introduces one-dimensional simplified models
that help to understand the above numerical results, and our
Conclusions are presented in Section \ref{sec:conclusions}.

\section{Reaction-advection-diffusion dynamics}
\label{sec:rad}

\subsection{General framework}
Let us consider $N$ interacting species with concentrations
$C_i(\br,t)$, $i=1,2,...N$, transported by a flow $\bv(\br,t)$
that we assume incompressible. The governing
reaction-advection-diffusion equations can be written as:
\BE
{\partial C_i \over \partial t}+\bv\cdot\nabla
C_i=\cf_i\left(C_1,...,C_N;k_1,...,k_M\right)+D_i\nabla^2C_i,
\label{raddim}
\EE
where the $\cf_i\left(C_1,...,C_N;k_1,...,k_M\right)$ describe
the interaction dynamics of excitable type  among the  components,
and the parameters $k_i$ are the reaction rates.
Although in some realizations of excitable reactions the diffusion
coefficients can vary widely from one species to the other, as for
example when it takes place in a gel medium, in liquid media
diffusion coefficients are rather similar and, for simplicity, we
take in the following $D_i =D$, $\forall i$. It is convenient to
adimensionalize Eq.~(\ref{raddim}) to have a clearer view of the
processes involved. To this end, let us identify typical scale,
$L$, and speed $U$ of the flow. A typical time scale is thus
$L/U$, and we perform the change of variables:
\BE
\br \rightarrow \br'\equiv\frac{\br}{L}, \ t \rightarrow
t'\equiv\frac{tU}{L}, \ \bv \rightarrow \bv'\equiv\frac{\bv}{U},
\label{xtvadimen}
\EE

 We assume that the
concentrations are already expressed in some convenient
dimensionless units, so that the reaction rates have units of
inverse time, and use one of the reaction rates, say $k_1\equiv
k$, to define dimensionless reaction terms:
\BA
\cf_i \rightarrow
\cg_i\left(C_1,...,C_N; \frac{k_2}{k},...,\frac{k_M}{k}\right)
&\equiv& \\  \nonumber
 k^{-1}\cf_i\left(C_1,...,C_N; k, k_2,...,k_M\right) &.&
\label{Fadimen}
\EA

With these changes, Eq.~(\ref{raddim}) reads
\BE
{\partial C_i\over \partial t}+\bv\cdot\nabla C_i=\Da~
\cg_i\left(C_1,...,C_N;\epsilon_2,...,\epsilon_M\right)+
\frac{1}{\Pe}\nabla^2C_i  \ \ \ ,
\label{rad}
\EE
where the primes have been omitted for notational simplicity. We
have scaled the reaction rates in terms of the first one
$\epsilon_i
\equiv k_i/k$, $i=2,3,...,M$, and
\BE
\Da \equiv \frac{kL}{U}\ {\rm and}\ \Pe \equiv \frac{LU}{D}
\label{numbers}
\EE
are the Damk\"{o}hler and the P\'{e}clet number, respectively. The
Damk\"{o}hler number measures the reaction speed in terms of the
advection, whereas $\Pe$ is the ratio of advection to diffusion at
scale $L$. The product $\Pe \Da$ measures the importance of
reaction with respect to diffusion. We will be interested in the
regime of large $\Pe$, so that diffusion is negligible except at
scales much smaller than system size, and explore a range of
values of $\Da$. We consider several two-dimensional models of
flow, and also one-dimensional simplifications of them. Sensible
comparisons of the behavior under different flows would be
facilitated by the introduction of the above adimensional
framework, although perfect correspondence can not be expected
when they are not dynamically similar.

\subsection{Reaction and flow models}
\label{subsec:models}
As a concrete example of reaction scheme of the excitable type, we
focus on the FitzHugh-Nagumo (FN) model \cite{excitable,Murray}.
We note however that we expect all our qualitative results to
apply to the general class of excitable systems. In fact, some
early results for closed flows \cite{PRL} have been already
checked for models of plankton dynamics\cite{GRL}. The FN model
consists in a dynamics of the type (\ref{rad}) for two interacting
species, of concentrations $C_1$ and $C_2$, and reaction terms:
\BA
\cg_1 &=& f(C_1)-C_2, \ \ f(C_1)=C_1(a-C_1)(C_1-1)
\label{FN1} \\
\cg_2 &=& \epsilon (C_1 -\gamma C_2)
\label{FN2}
\EA
$\epsilon$ ($=\epsilon_2$) is the ratio between the two time
scales. The FN model shows excitable behavior when $\epsilon
\ll 1$ so that there is a separation between the fast evolution of
the active component, $C_1$, and the slow evolution of $C_2$, the
passive one or {\sl inhibitor}.

In a homogeneous system, Eq. (\ref{rad}) becomes $\dot C_i = \Da
\cg_i$. When $C_2=0$, this dynamical system with Eq. (\ref{FN1})
describes dynamics of a bistable reaction, so that initial
conditions for $C_1$ below a threshold value $a$ decay towards the
{\sl unexcited} or {\sl rest} state $C_1=0$, whereas initial
conditions above $a$ evolve to the {\sl excited} state $C_1=1$ in
a time of the order of $\Da^{-1}$. But Eq. (\ref{FN2}) implies
that, as soon as $C_1$ grows above zero, the inhibitor $C_2$ grows
also (on a time scale a factor $\epsilon$ slower) and as a result
the excited state $C_1
\approx 1$ is only a transient: in a time of the order of
\BE
\tau_e \approx C_2^M(\epsilon \Da)^{-1} \ ,
\label{exctime}
\EE
$C_2$ reaches $C_2^M$, which is the local maximum value of the
function $f(C_1)$, and then deexcitation occurs. After a time of
the order of $(\epsilon
\Da \gamma)^{-1}$, during which the system can not be excited (the
{\sl refractory} state) $C_1$ and $C_2$ return back to the fixed
point or equilibrium value $C_1=C_2=0$. In the following we will
use the values $\epsilon=10^{-3}$, $\gamma=3.0$, and $a=0.25$ for
the parameters in the local FitzHugh-Nagumo dynamics. In this
case, $C_2^M\approx 0.1$.

 The flow is assumed to be imposed externally so that
the chemical dynamics has no influence on the velocity field,
${\bf v}({\bf r},t)$. We consider two different kinds of chaotic
flows which, in other contexts, are known to behave rather
differently: closed and open flows. In the first situation, fluid
particles remain in a bounded region of space, and the flow
produces mixing in the whole fluid. In the case of open flows,
fluid particles enter the system and typically, after some time,
they leave it. Interesting situations arise when, as an effect of
the stirring, there are special orbits never leaving the system.
For hyperbolic chaotic flows, such orbits form a fractal set of
zero measure, the {\sl chaotic saddle} \cite{open} with stable and
unstable manifolds. When a set of particles is released on that
flow, most of them leave the system rapidly. But those on
trajectories coming close to the stable manifold of the chaotic
saddle become attracted by it and remain longer in the system.
They finally escape the saddle, at a characteristic escape rate
$\kappa$, tracing closely its unstable manifold.

As a simple two-dimensional and incompressible velocity field,
we consider an archetype of closed chaotic flows, the
motion generated by two alternating sinusoidal shear flows
oriented along the $x$ and $y$ direction for the first and second
half of a flow period, $T$, respectively \cite{SinusFlow}:
\begin{eqnarray}
v_x(x,y,t) &=& {A \over T} \Theta
\Bigl({T \over 2} - t \bmod T \Bigr ) \sin{\left ({2 \pi y \over
L}+\phi_i\right)} \nonumber \\
v_y(x,y,t) &=& {A \over T} \Theta
\Bigl(t \bmod T - {T \over 2} \Bigr ) \sin{\left ({2 \pi x \over
L}+\phi_{i+1}\right)},
\label{closedflow}
\end{eqnarray}
where $\Theta$ is the Heaviside step function. Note that all the
geometric details of the stirring by the flow depend on the
parameter $A$, while $T$ sets the speed of stirring without
altering the trajectories of the fluid elements. In order to avoid
KAM tori acting as transport barriers, typically present in
time-periodic flows, the velocity field is made aperiodic by
introducing a random phase $\phi_i$, which takes on independent
values in each half period, and is uniformly distributed in the
range $[0,2\pi]$. The fluid is confined in a square of lateral
size $L$ with periodic boundary conditions, so that the flow is
closed. $L$ and $T$ fix space and time scales for the flow, and
$L/T$ is a typical velocity. Thus we can adimensionalize Eq.
(\ref{raddim}) in terms of these quantities, so that $\Da=kT$ and
$\Pe=L^2(DT)^{-1}$. This leads to Eq. (\ref{rad}), written in
units for which $L=T=1$.  We set the remaining adimensional
parameter $A/L=0.7$, for which the flow is nearly ergodic. The
numerically computed Lyapunov exponent is $\mu\approx 1.66/T$.

As a simple example of open flow, we take a blinking vortex-sink
system \cite{Aref89,KarolyiPR97} consisting of two alternately
opened point sinks in an unbounded two-dimensional domain. Around
each sink, the velocity field is a combination of a point-vortex
and a point-sink given by the complex potential
\BE
w(z)=-(Q+iK)\ln|z-z_s| \ .
\label{potential}
\EE
$z_s$ gives the position of the sink in the complex plane $\{z, z\in \mathcal{C}\}$,
so that $z-z_s=r e^{i\phi}$ defines polar coordinates $(r,\phi)$ around the
sink. The imaginary part of $w(z)$, $\Psi=-K \ln r -Q\phi$ is the streamfunction,
from which the fluid particle equations of motion are
\BA
\dot r &=& \frac{1}{r}{\partial\Psi\over\partial\phi} = -\frac{Q}{r} \nonumber \\
r \dot \phi &=& -{\partial\Psi\over\partial r} = \frac{K}{r} \ .
\label{motionsink}
\EA
  The fluid trajectories can be explicitly integrated:
\BA
r(t) &=& \sqrt{r_0^2-2 Q t}   \nonumber \\
\phi(t) &=& \phi_0-\frac{K}{Q}\ln \frac{r(t)}{r_0} \ .
\label{motionsinkint}
\EA
The fluid particles come from infinity following logarithmic
spirals of circulation given by $2\pi K$. The flow is
incompressible everywhere, except at the sink core $z_s$, where an
area of fluid $2\pi Q$ disappears per unit of time (the
trajectories in a circular region of this area around $z_s$ have
their trajectories undefined after one time unit because of
(\ref{motionsinkint}) and they should be understood to leave the
system): we have thus an open flow. The motion is not chaotic if
$z_s$ is a unique static point. The {\sl blinking vortex-sink}
flow consists in considering the active sink position to be at
$z_s=(0,d)$ during half a period $T/2$, and at $z_s=(0,-d)$ during
another half a period. This corresponds physically to opening and
closing alternatively two sinks separated by a distance $2d$.
Typical space, time, and velocity scales are $d$, $T$, and $d/T$,
respectively. From them, $\Da=kT$, and $\Pe=d^2(DT)^{-1}$, and we
can use Eq. (\ref{rad}) in the units in which $d=T=1$. The flow is
fully characterized by the adimensional sink strength
$\eta=QT/d^2$ and the ratio of vortex to sink strength $\xi=K/Q$.
We take $\eta=1$ and $\xi=10$. For these parameter values, the
Lyapunov exponent on the saddle and the escape rate from it
are\cite{KarolyiPR97} $\mu=2.19/T$ and $\kappa=0.54/T$,
respectively.

\section{Numerical results}
\label{sec:numerics}

The numerical integration of the reaction-advection-diffusion
problem has been carried out on a square grid of $1000 \times
1000$ points, with grid size $\Delta x$, by using a simple
semi-Lagrangian scheme for the transport processes combined with a
fourth-order Runge-Kutta method for the time integration of the
local chemical dynamics. The semi-Lagrangian advection step at
time $t$ consists in calculating, from any gridpoint, a
time-backwards Lagrangian trajectory for a time $\Delta t$. Then
the concentrations at this fluid element are calculated by
bilinear interpolation from the concentrations on the grid at time
$t-\Delta t$, and these concentration values are then assigned to
the starting gridpoint at time $t$. The interpolation step
introduces an effective diffusion $D\approx (\Delta x)^2/\Delta
t$, which limits the maximum $\Pe$ number we can attain.
Since the numerical diffusion is not uniform in space, we also
include an explicit diffusion step corresponding to the same $\Pe$
number.

Initially the system is in the homogeneous steady state,
$C_1=C_2=0.0$, and then it is perturbed by a localized Gaussian
pulse in the concentration of the {\it activator} component
\begin{equation}
C_1(x,y,t=0)=C_0 \exp{(-(x^2+y^2)/2 \l_0^2)},
\label{perturbation}
\end{equation}
where $C_0$ is chosen to be larger than the excitation threshold
$a=0.25$, and the size of the perturbation, $l_0$, is much
smaller than the system size, $L=1$. The dependence of the results
on the particular values of $C_0$ and $l_0$ will be discussed
later. The inhibitor component, $C_2$, is not perturbed initially.
We study the response of the system for different values of the
adimensional reaction rate $\Da$, keeping the rest of the
parameters fixed. In the absence of flow ($\Da=\infty$) the
initial condition (\ref{perturbation}) produces a circular ring of
excitation (a {\sl target} wave; in fact the structure of the
target wave is such that the excitation ring is followed by a
refractory ring) that expands in radius until reaching the system
boundaries. We will see that this behavior is strongly modified at
finite $\Da$.

\subsection{Closed flow}
The model (\ref{closedflow}) is integrated numerically on the unit
square, $L=1$, with periodic boundary conditions, from the initial
condition described above. The resolution is thus $\Delta
x=10^{-3}$. We use $\Delta t= 10^{-3}$ (in units of the flow
period) and thus $\Pe \approx 1000$.

\begin{center}
\begin{figure}
\epsfig{file=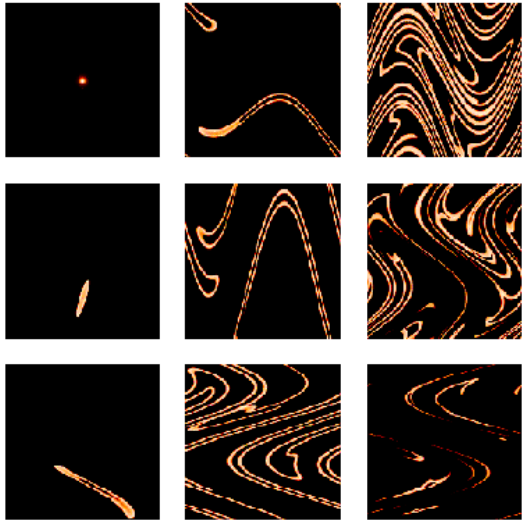,width=0.4\textwidth}
\caption{
Activator concentration at different times under stirring by the
closed flow, for $\Da=300$ and $\Pe\approx 1000$. The initial
condition had $C_0=0.5$ and $l_0=0.01$. The sequence runs from top
to bottom, and then from left to right. The total time lapse is 3
periods of the flow. Dark and clear gray level indicates,
respectively, low and high activator values. A noncoherent process
of global excitation is seen. }
\label{fig:noncoherent}

\end{figure}
\end{center}

\begin{center}
\begin{figure}
\epsfig{file=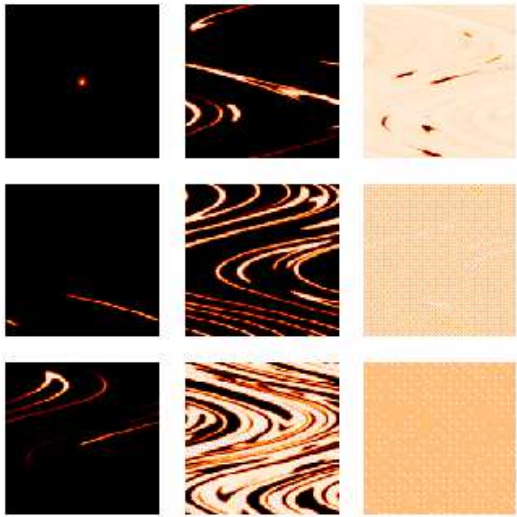,width=0.4\textwidth}
\caption{
Same as Fig. (\ref{fig:noncoherent}), but at $\Da=25$.
The interval between snapshots is of 1.2 time units.
A process of coherent global
excitation is seen. At longer times (not shown) the system returns
homogeneously to the unexcited state. }
\label{fig:coherent}
\end{figure}
\end{center}

Snapshots of the spatial structure for $\Da=300$ and $\Da=25$ are
shown in Figs. \ref{fig:noncoherent} and \ref{fig:coherent},
respectively. In the first case, the localized perturbation gives
rise to an excited patch (with its interior in the refractory
state) that is the stirred version of the circular wave front that
would be produced in the absence of flow. The patch is elongated
into a convoluted filamental structure by the chaotic flow and
eventually visit all the points of the system. The filaments have
a characteristic {\sl double-line} structure with refractory area
in the center.
The excitation is global, in the sense that all the
points of the system have become excited at some moment, but is
not coherent, since only a part of the system is excited at a
given moment, being the rest in the refractory or in the
equilibrium state.

In a range of smaller $\Da$ numbers, a qualitatively different
phenomenon occurs: The initial patch is again stretched into a
growing filament, but now the filaments are thinner that 
prevents the formation of refractory region within them.
This results in a
coherent global excitation when the filaments fill up the whole system. Fig.
\ref{fig:coherent} is a representative example of this situation.
Once fully excited, the system remains homogeneous and its
subsequent decay to the unexcited state occurs everywhere at the
same time. In this second part of the dynamics, mixing becomes
irrelevant since there is nothing to mix in an homogeneous
configuration. 

At still smaller $\Da$ (faster stirring or slower chemistry), a
sharp transition to a new dynamic regime occurs: below a critical
$\Da$ number, $\Da_c$ the excitation dies without propagating
significantly: dilution is fast and dominates over the growth
rate of the activator. By {\sl dilution} we mean the process of
mixing of the excited patch with the surrounding unexcited fluid,
leading to the decreasing and elimination of excitation in the
patch. This mixing is originated by the diffusive flux from inside
the patch (high value of the activator) to outside (low activator
value). 
The value of $\Da_c$ depends only slightly on
the details of the initial condition (as long as $C_0$ remains
suprathreshold and $l_0$ much smaller than system size and above
some diffusion-controlled minimum size, of the order of Eq.
(\ref{wu}), discussed below). For example, if $C_0=1$, and
$l_0=0.1$, $0.05$, and $0.02$, one finds $\Da_c
\approx 12.0$, $14.2$, and $16.3$, respectively.

\begin{center}
\begin{figure}
\epsfig{file=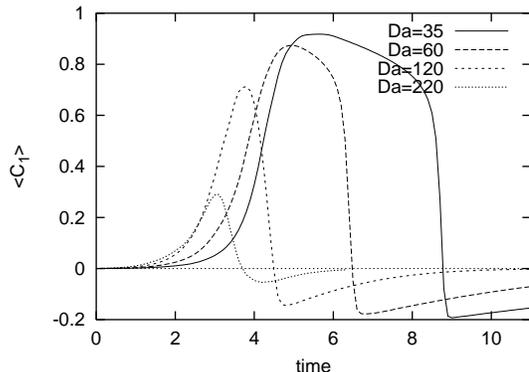,width=0.4\textwidth}
\caption{Time evolution of the mean concentration at various values of $\Da$.
$C_0=0.5$ and $l_0=0.01$. Time is in units of $T$. By further
decreasing $\Da$, this time evolution changes suddenly, at
$\Da_c$, to a fast decay of the initial condition that is
indistinguishable, at the scale of this plot, from the horizontal
axis at zero mean concentration.}
\label{fig:C_vs_t}
\end{figure}
\end{center}

\begin{center}
\begin{figure}
\epsfig{file=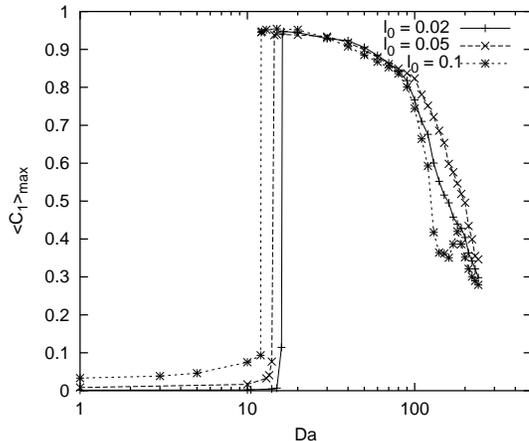,width=0.4\textwidth}
\caption{The maximum value attained by $<C_1>$ as a function of
$\Da$. In all cases the initial condition had $C_0=1$, and $l_0$
was as indicated. }
\label{fig:Cmax_vs_Da}
\end{figure}
\end{center}

Figs. \ref{fig:C_vs_t} and \ref{fig:Cmax_vs_Da} summarize the
different situations. Figure \ref{fig:C_vs_t} shows the time
dependence of $\left<C_1\right>$, the space-average of the
concentration of $C_1$, and Fig.\ref{fig:Cmax_vs_Da}, the maximum
value attained by $\left<C_1\right>$. For increasing $\Da$ the
coherence of the global excitation is gradually lost so that the
maximum value of the average concentration is below the one
corresponding to the fully excited state.

\subsection{Open flow}

Since an infinite domain can not be simulated easily in the
computer, a square domain of size $L/d=6$ is considered instead.
With a lateral discretization of 1000 points, this leads to
$\Delta x=6\times 10^{-3}$. We use $\Delta t = 2 \times 10^{-3}$
(in units of the flow period) and thus $\Pe \approx 56$.
Concentrations at the boundaries are kept at the fixed-point
values $C_1=C_2=0$. The interior of the domain is initialized also
in this state except for the perturbation in $C_1$, Eq.
(\ref{perturbation}), located in the middle position between the
two sinks.

\begin{center}
\begin{figure}
\epsfig{file=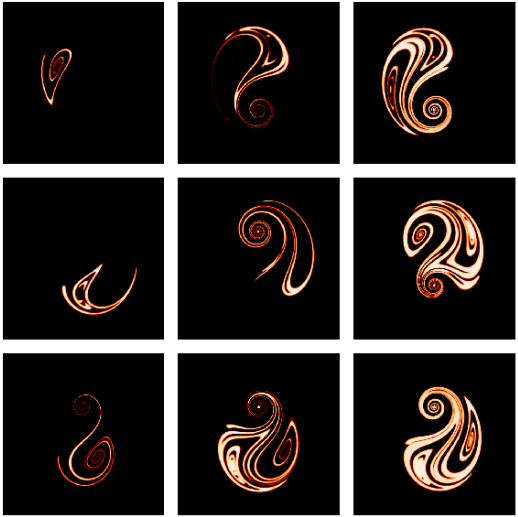, width=.4\textwidth}
\caption{Time evolution of the activator concentration under the open flow.
The snapshots are shown every 0.7 time units, from top to bottom
and then from left to right, starting 0.7 time units after the
initial perturbation, $\Da=50$ 
and $\Pe\approx 56$. After the last time shown,
excitation is maintained indefinitely in the system, and the
long-time pattern, which follows the shape of the unstable
manifold of the chaotic saddle, repeats periodically in synchrony
with the flow. }
\label{fig:open75}
\end{figure}
\end{center}

\begin{center}
\begin{figure}
\epsfig{file=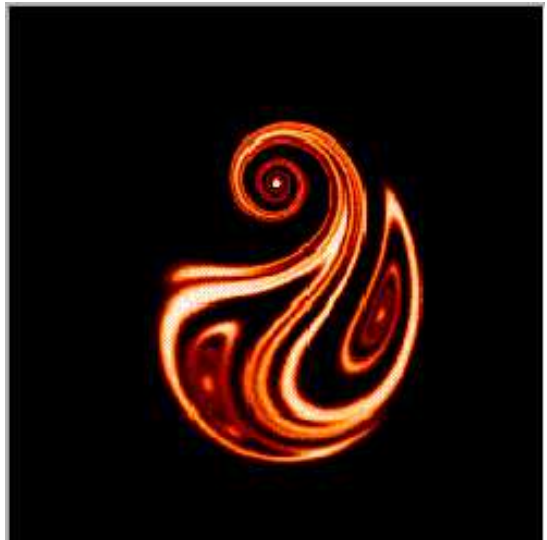, width=.4\textwidth}
\caption{snapshot of the activator concentration maintained
in the system at long times, for $\Da=50$.}
\label{fig:openDaxx}
\end{figure}
\end{center}

For small $\Da$, as in the closed flow case, the perturbation is
diluted by the flow before significant wave propagation occurs,
and the excited material leaves soon the system through one of the
sinks. By increasing $\Da$, a sharp transition to a new regime
occurs: In response to the persistent arrival of unexcited
reactants from the boundaries, and despite the continuous loss of
fluid through the sink, a steady pattern of excitation is
permanently sustained by the autocatalytic behavior of the
activator. An example of the time evolution is shown in Fig.
\ref{fig:open75}, and a snapshot in Fig. \ref{fig:openDaxx}. The
excited pattern closely traces a fattened version of the unstable
manifold associated to the chaotic saddle of this open
flow\cite{KarolyiPR97}, and as this manifold, it fluctuates
periodically in time. The value $\Da_c
\approx 14.5$ at which this transition occurs 
is in the range of the values
obtained for the closed flow, thus suggesting that the mechanism
for it is rather local and independent of the details of the flow.

The filaments in the excited pattern fatten up with increasing
$\Da$. 
Suddenly, a new regime is reached above a second critical
$\Da\approx 90.5$: the excitation initially accumulates at the
unstable manifold of the chaotic saddle, as before, but this is
just a transient that is followed by an irregular recovery of the
equilibrium ($C_1=C_2=0$) state everywhere. This new regime has
some analogies with the large $\Da$ behavior under the closed
flow, for which a noncoherent excitation occurred (and the
transition value of $\Da$ is of the same order), but here it
appears suddenly as a function of $\Da$, and the excitation does
not visit the full system but remains close to the unstable
manifold of the chaotic saddle.

\begin{center}
\begin{figure}
\epsfig{file=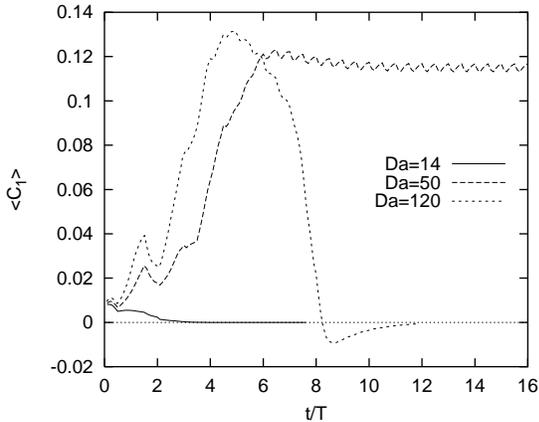, width=.4\textwidth}
\caption{Time evolution of the average concentration $<C_1>$ for three values
of $\Da$. $C_0=0.5$, $l_0=0.05$. 
}
\label{fig:openC1_vs_t}
\end{figure}
\end{center}

\begin{center}
\begin{figure}
\epsfig{file=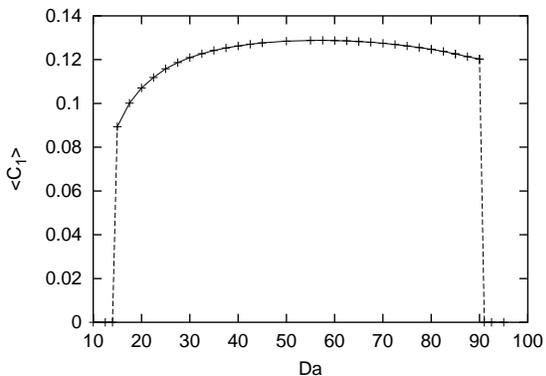, width=.4\textwidth}
\caption{Long-time asymptotic value of the average concentration $<C_1>$ as
a function of $\Da$. It is non-zero in a
range of intermediate values.}
\label{fig:openTotalC1nueva}
\end{figure}
\end{center}

The whole behavior is summarized in Figs. \ref{fig:openC1_vs_t}
and \ref{fig:openTotalC1nueva}, where the time evolution of the
mean value of the activator, and its asymptotic long time value
(not the maximum value as in Fig. \ref{fig:Cmax_vs_Da}), is
plotted versus $\Da$. There is a range of $\Da$ in which a finite
amount of excited fluid remains permanently in the system, despite
the openness of the flow.


The existence of critical values of $\Da$ or equivalently, of
critical stirring rates can be seen to be a consequence of the
competition between a number of processes. In the closed flow
case, advection and diffusion tends to homogenize and dilute the
excited patch, while the excitable dynamics increases the local
concentration of the active component wherever the excitation
threshold is exceeded. In the open flow, there is an additional
factor: the escape rate of fluid particles from the system. In
order to gain further insight we attempt to separate the essential
ingredients that contribute to the observed behaviour, and
consider reduced models of the problem.

\section{Reduced models}
\label{sec:reduced}
\subsection{One-dimensional baker model for the closed flow}
\label{subsec:bakerclosed}

The main effect of chaotic advection is to stretch and fold fluid
elements producing the filamentary patterns visible in the figures
of the last Section. Perhaps the simplest model of this is the so
called {\it baker transformation}. A single action of the baker
transformation on the unit square can be described as a stretching
along the $y$ axis by a factor of two, followed by compression by
a factor of two along the $x$ axis. Then the resulting rectangle
is cut into two pieces of unit length along the $y$ direction and
placed back on the unit square. This model of chaotic advection
neglects spatial non-uniformities of the stretching and curvature
of the filaments, present in a general flow. Nevertheless, since
it is a discrete-time map, it is strongly non-uniform in time.

If the initial perturbation is taken to be homogeneous in the $y$
direction this is preserved by the baker transformation and the
problem becomes one-dimensional. Even for arbitrary initial
conditions the concentrations are rapidly homogenized along the
$y$ direction by the repeated stretching, and after a short time
the one-dimensional description becomes relevant. This may be
regarded as a general feature of transport problems in the
presence of stirring, since one can associate local stretching
directions to any point of the flow and the problem can be reduced
to the description of the filamental structure in the transverse
direction. In the one-dimensional formulation the baker
transformation ${\cal T}$ acts by replacing the concentration
field by two copies compressed by a factor of two placed next to
each other. To better represent the process of filament folding,
the left half is not a copy but the mirror image of the right
half.
\BA
{\cal T}: \ \ \ \ \ \ x &\to& {\cal T}x = {x/2, (2-x)/2}.
\nonumber \\
          C_i(x) &\to& {\cal T}C_i(x) = C_i({\cal T}^{-1}x)
\label{baker}
\EA
Numerical simulations on the unit interval with periodic boundary
conditions of the one-dimensional FitzHugh-Nagumo system with the
baker transformation applied at discrete times ($t=nT, n=1,2,..$),
and diffusion and chemistry acting between them, show
qualitatively similar regimes to the two-dimensional closed flow
presented above, including the transition to global excitation
that occurs in an intermediate range of $\Da=kT$.
\begin{center}
\begin{figure}
\end{figure}
\end{center}

\begin{center}
\begin{figure}
\end{figure}

\end{center}
\begin{center}
\begin{figure}
\end{figure}

\end{center}
\begin{center}
\begin{figure}
\epsfig{file=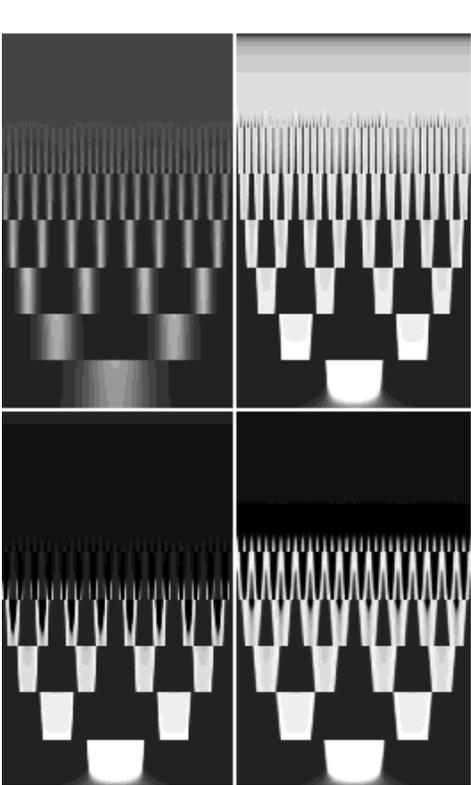, width=.35\textwidth}
\caption{
Spatiotemporal evolution of $C_1$ for $\Da=1$ and $\Pe=1000$
(upper left), $\Da=10$ and $\Pe=1000$ (upper right), $\Da=40$ and
$\Pe=1000$ (lower left), and $\Da=50$, $\Pe=400$ (lower right)
under the baker model. Space is in the horizontal direction, and
time runs in the vertical from bottom to top. Darker grey
represents smaller values of $C_1$. The discontinuities appear at
each application of the baker map, i.e., at times $T$, $2T$, $3T$,
etc. }
\label{fig:bakerClosed}
\end{figure}
\end{center}

Time evolutions of the activator spatial structure are shown in
Fig. \ref{fig:bakerClosed}, for three different values of $\Da$,
and two of $\Pe=L^2/DT$. The excited regions can be interpreted as
transverse cuts through filaments. The number of filaments is
doubled by each action of the baker map, while the decreasing of
their width may be, or may be not, compensated by the effect of
excitable growth. When growth is slow (upper left panel in Fig.
\ref{fig:bakerClosed}), the filaments become narrower until a
point in which diffusive mixing with the surrounding unexcited
fluid destroys them and excitation disappears. By increasing $\Da$
(upper right), the filament width reaches a minimum nonvanishing
value with its central $C_1$ concentration value well above the
threshold $a$. The effect of the baker map is here to join
together a number of filaments until diffusion homogenises the
distribution. Since the homogenized value of $C_1$ is above the
threshold, a coherent excitation follows. After some time the
excitation disappears homogeneously and the system returns to the
nonexcited state.

In the case of large $\Da$ (lower panels), the reaction is fast
enough to approach the refractory state in the middle of the
filaments before significant compression. The filaments thus
acquire the double-hump structure that was also seen in the full
two-dimensional simulation (Fig. \ref{fig:noncoherent}). This fact
works against the possibility of a coherent excitation, and there
are two mechanisms by which the noncoherent excitation dies. At
large enough $\Pe$ (lower left panel of Fig.
\ref{fig:bakerClosed}), the excited parts of the filaments are
narrow, and the periodic contraction produced by the baker map
eventually bring them below a width such that diffusion can
eliminate them, in a way similar to what happens with the full
filament at small $\Da$ (but here there is around abundant
refractory material that helps the process). When $\Pe$ is
decreased, the excited parts of the filaments travel faster, and
collisions leading to filament annihilation (because of the
refractory material arriving after the excited filament) are the
main mechanism killing the excitation (lower right panel of Fig.
\ref{fig:bakerClosed}).

The phenomenology found here is fully consistent with the
numerical simulations of Section \ref{sec:numerics}. But now, in
addition to having a much simpler numerics, the mechanisms are
easier to identify. Thus, stretching and folding, the
characteristics implemented in the baker map, are enough to
understand the effects of chaotic advection on excitable
advection-reaction system. We stress however that the $\Da$ values
at which the different transitions occur are of the same order,
but not identical, to the ones found in the two-dimensional
models. This was expected since there is no complete dynamical
similarity between the present baker model and the flow models of
Sect. \ref{sec:numerics}.

In the baker map model the spatial
structure is, by construction, periodic with period $L/n,
n=[t/T]$. Thus the evolution of the system can be fully described
by solving the same problem on an interval compressed by a factor
of two at times $t=nT$ with periodic boundary conditions. This
suggests, in general, that the evolution of the system can be
captured by focusing on the transverse profile of a single
filament subject to a typical stretching, and taking into account
the decreasing separation between the filaments by appropriate
boundary conditions. In fact, the main mechanism controlling the
final homogenized value is the competition between the
compression by the flow, and the tendency to expansion due to
reaction-diffusion. This mechanism is better analyzed by
considering an isolated single filament, as will be done in
Subsection \ref{subsec:filament}.

\subsection{One-dimensional baker model for the open flow}
\label{subsec:bakeropen}

As in the closed flow case, we can implement the essentials of
chaotic advection in open flows: stretching, folding, and escape,
by a one-dimensional version of the open baker map. At times
$nT$, $n=1,2,...$, the unit interval ($L=1$) is compressed a
factor of three; two copies of the resulting compressed
configuration are placed back into the initial square (one of them
with orientation reversed) and the remaining third is filled with
unexcited material ($C_1=C_2=0$). This represents the loss of one
third of the fluid per map step, and its substitution by fresh
reactants. Standard diffusion with periodic boundary conditions,
and FitzHugh-Nagumo dynamics act between successive applications
of the map.

\begin{center}
\begin{figure}
\epsfig{file=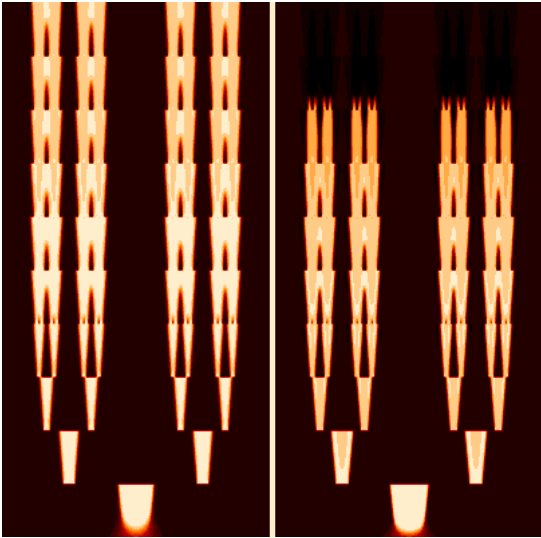,width=.4\textwidth}
\caption{Concentration of $C_1$ under the open baker map for $\Da=30$ (left)
and $\Da=50$ (right). Space is in the horizontal direction, and
time in the vertical, running from bottom to top. Darker grey
levels correspond to lower values of $C_1$. $\Pe=10^4$. }
\label{fig:bakerOpen}
\end{figure}
\end{center}

The phenomenology observed is again qualitatively consistent with
the two-dimensional simulations. For small $\Da$ the initial
excitation is diluted before significant propagation. At larger
$\Da$ (Fig. \ref{fig:bakerOpen}, left panel ), the excitation
approaches the chaotic saddle of this map, which is a standard
Cantor set, and covers it with a finite width. A dynamic
equilibrium is reached between filament merging and filament
replication, so that the excitation is maintained indefinitely in
the system. Increasing further $\Da$ leads to a second transition
to a situation in which the excitation finally disappears, in much
the same way as in the closed flow. Again this happens when the
filaments begin to develop the refractory state in its interior,
and it may occur by two mechanisms: the one shown in the right
panel of Fig. \ref{fig:bakerOpen} which involves complex filament
interaction, or simply the repeated contraction of the narrow
excited parts at both sides of the refractory center. This last
mechanism dominates at very large $\Pe$.

Both in the open and in the closed flow case, the mechanisms
leading to transitions and qualitative changes in the excitation
behavior seem to be linked to properties of individual filaments,
namely, the existence of a minimum width below which the filament
disappears, and the development of a double-hump shape by
increasing $\Da$. Both phenomena can be understood to great detail
by focusing on the behavior of an isolated filament.

\subsection{A one-dimensional filament model}
\label{subsec:filament}

We can address the analysis of the one-filament problem by
replacing the flow by a time-continuous stretching in a pure
strain flow, $v_x = -\lambda x, v_y= \lambda y$. This has in
common with chaotic advection and with the baker map the local
contraction and expansion along special directions, although it
misses completely the folding behavior that leads to filament
interaction at long times. According to the discussion above the
relevant dynamics is along the convergent direction ($x$) of the
flow. Thus, we propose that the evolution of concentrations
$C_i(x,t)$ in a chaotically advected excitable medium can be
described locally by
\begin{equation}
{\partial \over \partial t}{C_i}-\lambda x {\partial \over
\partial x} {C_i} = \cf_i(C_1,...,C_N) + D {\partial \over
{\partial x}^2} {C_i},
\label{strain}
\end{equation}
where $\lambda$ is the strain due to the
flow\cite{Martin2000,PRL,ZoltanFocus}. In general, the strength
and direction of the stretching fluctuates in space and time. Thus
a suitable prescription for fixing a unique $\lambda$ should be
established. This issue will be discussed later. A way to take
into account multifilament situations in the framework of Eq.
(\ref{strain}), is to impose periodic boundary conditions on an
exponentially shrinking interval of length $L=\exp{(-\lambda t)}$,
taking into account the decreasing interfilamental distance. But
we will consider here the case of an isolated filament in a large
(ideally infinite) one-dimensional domain.

We note that the one-dimensional model (\ref{strain}) does not
conserve the amount of fluid on the line. This is more clearly
seen be rewriting it as
\BA
{\partial \over \partial t}{C_i} &+& {\partial \over
\partial x}\left(-\lambda x  {C_i} -  D {\partial \over {\partial x}}
{C_i}\right) = \\ \nonumber & &\cf_i(C_1,...,C_N) - \lambda C_i
\ .
\label{strain2}
\EA
Whereas the left hand side is clearly written in a
flux-conservative form, the term $-\lambda C_i$ in the right hand
side represents fluid escape from the line at a rate $\lambda$.
The reason for that is that Eq. (\ref{strain}) comes from a strain
flow in which there is motion along the $y$ axis, and the term
$\lambda C_i$ is simply the flux in that direction for
concentrations homogeneous along $y$: $\lambda C_i(x) =
\partial_y(\lambda y C_i(x))$.

The pure strain flow has a time scale, $\lambda^{-1}$, that can be
used to adimensionalize times, but there is no typical length scale.
However, we
can measure lengths in units of the diffusion length
$\sqrt{D/\lambda}$ and then Eq. (\ref{strain}) becomes
\BE
{\partial \over \partial \bar t}{C_i}- \bar x {\partial \over
\partial \bar x} {C_i} = \widetilde{\Da} \cg_i(C_1,...,C_N) +
{\partial^2 \over {\partial \bar x}^2} {C_i},
\label{strain3}
\EE
with $\widetilde{\Da}=k/\lambda$, $\bar t=\lambda t$, and $\bar x
= x (D/\lambda)^{-1/2}$. Thus, we can always set
$\Pe=L^2\lambda/D=1$ by choosing the units of $L$ or, in other
words, the only effect of variations of diffusion strength in this
model is a change in spatial scale. Qualitative changes can only
occur by varying $\widetilde{\Da}$, that is, by changing $k$ or
$\lambda$. This is clearly a limitation of the model and tell us
that it can only be trusted in regions where there are well
separated filaments of size much smaller than characteristic
spatial scales of the velocity field (which are neglected when
assuming a pure strain). We expect this to be a reasonable global
approximation at sufficiently large $\Pe$. Other phenomena
neglected by this one-dimensional model are strain inhomogeneities
and departure from one-dimensionality.

In this Section we mainly present our results in terms of the
parameter $\widetilde{\Da}$ of Eq.~(\ref{strain3}), but eventually
we would need to return to the units of Eq.~(\ref{strain}), where
the individual processes and scales are more easily identified. In
the search of clarity, quantities representing lengths will be
marked with an overbar when measured in the units of Eq.
(\ref{strain3}), that is, in units of the diffusion length. We
note also that changing the strain $\lambda$ in Eq. (\ref{strain})
changes $\widetilde{\Da}=k/\lambda$ and also the units of space
and time in Eq. (\ref{strain3}).

\begin{center}
\begin{figure}
\epsfig{file=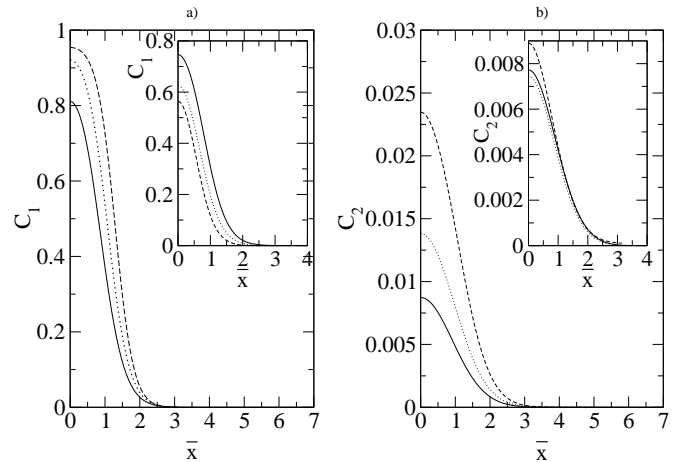,width=.40\textwidth,angle=-90}
\caption{a) $C_1$ profiles for stable one-humped solutions of Eq. (\ref{strain3}).
In the inset the corresponding unstable solutions. Solid line is
for $\widetilde{\Da}=12.99$, close to the disappearance of the
filament solution, dotted-line for $\widetilde{\Da}=16.77$ and
dashed-line for $\widetilde{\Da}=25$. b) The same as a) but for
the $C_2$ concentration field. All the curves are symmetric with
respect to $\bar x=0$, so that only positive $\bar x$ values are
plotted.}
\label{fig:1hfilament}
\end{figure}
\end{center}

Numerical solution of Eq. (\ref{strain3}) for $\widetilde{\Da}$
not too small reveals that its long-time attractors are steady
pulses of excitation concentrated near the origin. They can be
interpreted as transverse cuts of the filaments observed in the
two-dimensional models. Examples are shown in Figs.
\ref{fig:1hfilament}a) and \ref{fig:1hfilament}b), where we plot
the $C_1$ and $C_2$ concentration fields, respectively, for
different values of $\widetilde{\Da}$ (the insets will be
discussed below).
  The steady finite width of the filaments arises from compensation
between the contracting tendency of the strain, and the expanding
tendency of the combined effect of diffusion and reaction. A
simple quantitative argument \cite{PRL} formalizing this consists
in identifying the equilibrium half-width of the filament
solutions of (\ref{strain}), $w_s$, as the distance to the center
at which the strain speed $\lambda w_s$ exactly compensates the
speed the front would have in the absence of strain, $v_f$. Since
this velocity may be calculated for small $\epsilon$
\cite{Murray}: $v_f = (1-2a)\sqrt{D k/2}$, $w_s$ is obtained from
$\lambda w_s=v_f$. In the adimensional space units and parameters
of Eq. (\ref{strain3}) it reads:
\BE \bar w_s=(1-2a)\sqrt{\widetilde{\Da} \over 2} \ .
\label{ws}
\EE

The existence, for  $\widetilde{\Da}$ above (or $\lambda$ below) a
given value, of these steady filaments with finite width provides
an explanation for most of the phenomenology discussed in the
previous sections. In two-dimensional situations, the filament
will maintain its transverse shape while it expands in the
longitudinal direction. After repeated folding, it will cover the
whole system in the closed flow case, or cover the unstable
manifold of the chaotic saddle in the open flow case. What will
happen latter will depend on the interactions between different
parts of the excited filament, or on its response to strain
fluctuations.

\begin{center}
\begin{figure}
\epsfig{file=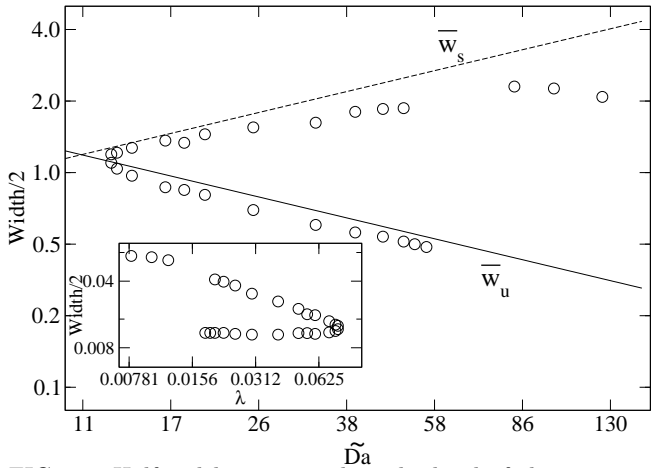,width=.40\textwidth,angle=-90}
\caption{Half-widths measured at the level of the excitation threshold
($C_1=0.25$) for the $C_1$ stable and unstable filament solutions
of Eq. (\ref{strain3}) in terms of $\widetilde{\Da}$ (log-log
scale). Circles show the numerical values and the lines the
analytical curves: solid-line from Eq.~(\ref{wu}) and dashed-line
from Eq.~(\ref{ws}). The inset shows the same numerical widths but
in the dimensional units of Eq. (\ref{strain}) (we use $k=1$ and
$D=10^{-5}$) as a function of $\lambda$, to stress the insensivity
of the width of the lower (unstable) branch to the strain
$\lambda$.}
\label{fig:anchura}
\end{figure}
\end{center}

We observe that the steady-filament stable solution disappears for
$\widetilde{\Da} < \widetilde{\Da}_c \approx 12.5$). This provides
an explanation for the absence of excitation in the
two-dimensional simulations below a critical $\Da$: as far as the
results of the one-dimensional model can be extrapolated there, a
growing filament state can not be reached at small $\Da$ because
a steady (non-decaying) solution of the filament profile does not exist. 
To better understand the disappearance of the
filament solution, we note that, in addition to direct numerical
simulation, an alternative way of finding the steady filaments is
to solve by a shooting method \cite{NumRecipes} the steady state
version of Eq. (\ref{strain3}) which is obtained by setting
$\partial_{\bar t} C_i=0$. With this method one can obtain all the
steady solutions, not only those that are dynamically stable.
 It turns out that, in addition to the excited filament (and to
the trivial homogeneous solution $C_1(\bar x)=C_2(\bar x)=0$) ,
there is another pulse solution, which is dynamically unstable.
This unstable solution is shown, for several values of
$\widetilde{\Da}$, in the insets of Fig. \ref{fig:1hfilament}. It
contains a marginal amount of excitation, in the sense that
initial conditions with slightly less excited material evolve
towards the stable homogeneous state, and initial conditions
slightly more excited lead to the stable excited filament. It
represents the unstable point in function space at which the
activator growth exactly compensates the diffusive flux towards
the exterior. In Fig. \ref{fig:anchura} we plot the width of the
stable and unstable steady filament solutions of
Eq.~(\ref{strain3}). Here it is clear that the disappearance of
the stable filament arises from collision with the unstable pulse
in a saddle-node bifurcation. Physically, increasing strain
reduces the width of the stable filament, so that it approaches
the unstable one, which is the limit below which excitation
decays. The saddle-node bifurcation will occur when both widths
are equal.

\begin{center}
\begin{figure}
\epsfig{file=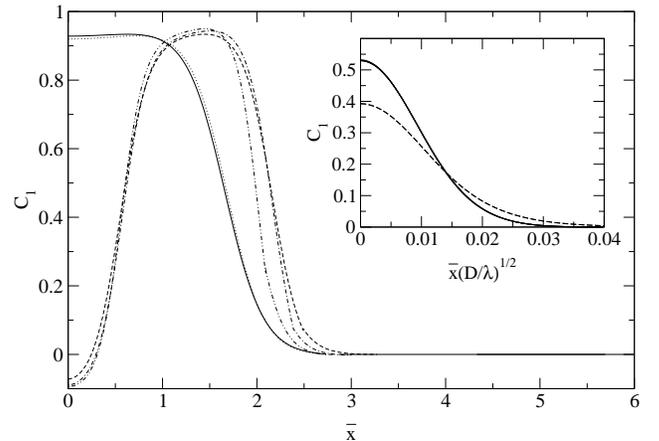,width=.40\textwidth,angle=-90}
\caption{Two-humped filament solutions for the $C_1$ field.
Solid-line is for $\widetilde{\Da}=45.45$, when the two humps
begin to develop. Dotted-line $\widetilde{\Da}=50$, 
dashed-line
$\widetilde{\Da}=83.33$, 
dashed-dotted line $\widetilde{\Da}=100$
and dashed-double-dotted line for $\widetilde{\Da}=125$. In the
inset we show the unstable solutions, for $\widetilde{\Da}=45.45$,
$50$, $52.63$, and $55.55$. In the scale of the plot, all of them
collapse into the same solid-line curve, when plotted in terms of
the physical space units of Eq. (\ref{strain}), as done here (we
use $k=1$ and $D=10^{-5}$). The dashed-line corresponds to the
analytical solution given by Eq.~(\ref{unstablepulse}). }
\label{fig:2hfilament}
\end{figure}
\end{center}

 Since the width of the unstable pulse around $\bar x=0$ is rather small, we
expect strain effects to be of minor importance in determining its
shape, at least when $\widetilde{\Da}$ is not too close to
$\widetilde{\Da}_c$. This is confirmed by the inset in
Fig.~\ref{fig:anchura}, and also in the inset of
Fig.~\ref{fig:2hfilament} (to be discussed later). Hence, we can
analytically estimate the shape of the unstable pulse and
$\widetilde{\Da}_c$ in the following way: Since the amount of
inhibitor is small everywhere for this solution (see the inset in
Fig. \ref{fig:1hfilament}b)), and since we are interested in the
situation $\epsilon \rightarrow 0$, we can approximate Eq.
(\ref{strain}) (for $\lambda \approx 0$) by
\BE
k C_1(a-C_1)(C_1-1)+D\frac{\partial^2}{\partial x^2}C_1 = 0
\label{steadybistable}
\EE
The unstable pulse, $C_1^{u}(x)$,
is the solution homoclinic to $C_1=0$. After
multiplication of (\ref{steadybistable}) by $\partial_x C_1(x)$,
integrations with respect to $x$, and application of the proper
boundary conditions, one finds \cite{saxena}
\BE
C_1^{u}(x)= C_+ - { C_+ - C_- \over 1-\frac{C_-}{C_+}\tanh^2
\left(
\frac{x}{w_u} \right)}
\label{unstablepulse}
\EE
with
\BE
C_\pm = \frac{2}{3}(1+a) \pm \sqrt{\frac{4}{9}(1+a)^2 - 2a} \ .
\label{Cpm}
\EE
$C_-$ is the maximum concentration at the center of the pulse, and
its half-width is given by $w_u = 2\sqrt{D/(a k)}$, or in the
adimensional units of Eq. (\ref{strain3}):
\BE
\bar w_u={2 \over \sqrt{a\widetilde{\Da}}} ,
\label{wu}
\EE
In the inset of Fig.~\ref{fig:2hfilament} we compare the
analytical curve from Eq.~(\ref{unstablepulse}) with the numerical
values. We see that they are similar but not identical. We have
checked that the reason for the discrepancy is the finite value of
$\epsilon$ ($\epsilon =10^{-3}$). We have also performed
calculations with smaller values of $\epsilon$ and seen that for
the values of $\widetilde{\Da}$ in the figure, the approximate
analytical solution (\ref{unstablepulse}) and the numerical one
become virtually identical when $\epsilon
\lesssim 10^{-5}$.

In any case, since the widths of analytical and numerical unstable
pulses are very similar, we can estimate $\widetilde{\Da}_c$ by
equating the above expressions, (\ref{ws}) and (\ref{wu}), found
for them: $\bar w_s=\bar w_u$, with the result
\BE
\widetilde{\Da}_c=\frac{2\sqrt{2}}{(1-2a)\sqrt{a}}
\label{Dac}
\EE
For $a=0.25$ this gives $\widetilde{\Da}_c \approx 11.31$, that
compares well with the numerically obtained value
$\widetilde{\Da}_c \approx 12.5$ (See Fig. (\ref{fig:anchura})).

The saddle-node disappearance of the filament solutions in this
one-dimensional filament model clearly gives an explanation for
the sudden disappearance of excitation propagation at small $\Da$
in the two-dimensional models discussed in Sect.
\ref{sec:numerics}, and in baker-like models. To make the
connection more quantitative, within the uncertainty given by the
observed weak dependence of $\Da_c$ on characteristics of the
initial perturbation (Fig. (\ref{fig:Cmax_vs_Da})), one needs to
identify the {\sl effective} strain $\lambda$ in Eq.
(\ref{strain}). If one identifies $\lambda
\approx T^{-1}$, which is a reasonable measure of the strain in
the models of Sect. \ref{sec:numerics}, then one has
$\Da=\widetilde{\Da}$ and finds good quantitative agreement
between the critical values of the Damk\"{o}hler number for the
filament model and the full two-dimensional simulations both in
the open and in the closed flow case. But it should be said that,
since the filaments are being advected by the flow, a more
consistent choice for $\lambda$ would be the Lagrangian mean
strain, given by the Lyapunov exponent $\mu$ of the advection
dynamics. In the open flow case, the Lyapunov exponent on the
chaotic saddle would be the analogous choice. These elections have
been shown to be quantitatively successful in other situations
\cite{ZoltanFocus}. With the values of $\mu$ stated in Sect.
\ref{subsec:models}, this leads to $\Da = 1.66
\widetilde{\Da}$ for the closed flow and $\Da = 2.19
\widetilde{\Da}$ in the open case. Now the agreement has deteriorated.
The effect of strain inhomogeneities may be rather important when
the filaments are wide and have some diffusive motion, since then
they can feel effective strains different from the Lagrangian one
corresponding to a fluid particle at its center. Other effects
related to the reduced dimensionality are also at play, since
quantitative departures from the two-dimensional simulations
appear already for the baker-like models of Sects.
\ref{subsec:bakerclosed} and \ref{subsec:bakeropen}. Thus, one
concludes that the one-dimensional filament model needs to be
improved to provide systematic quantitative predictions on the
behavior of reacting systems, but it does a very good job in
identifying the basic mechanisms and qualitatively modeling them.

Still remaining to be discussed within the framework of this
Section are the qualitative changes of behavior occurring in the
two-dimensional simulations at large $\Da$: the progressive loss
of coherence in the closed flow case, and the sudden disappearance
of the persistent pattern in the open flow case. These phenomena
were more or less coincident with the appearance of a double hump
structure in the filaments. Fig.~\ref{fig:2hfilament} shows that
indeed the stable filament solutions of the one-dimensional model
(\ref{strain3}) develop a double-humped structure for
$\widetilde{\Da} \gtrsim 45$. One can understand this by noticing
that the front solutions of Eq. (\ref{strain}) for $\lambda
\approx 0$ have a finite width limited by the time during which
excitation persists in the fluid particles (\ref{exctime}), i.e.,
$\tau_e = C_2^M(\epsilon\widetilde{\Da})^{-1}$, in units of
$\lambda^{-1}$. The width of the front is given in first
approximation by $w_f
\approx v_f\tau_e$. Interaction with the back of the front changes
$C_2^M$ to a smaller value $C_2^J$, which is the solution of an
algebraic equation\cite{Murray}. For $a=0.25$, $C_2^J\approx
0.067$. It is reasonable to expect that, when the total width of
the filament, $2w_s$, exceeds twice the width of the front $2
w_f$, the filament will become unexcited in the middle. This
argument would need corrections by the strain influence on $w_f$
and by the fact that the strain velocity in the middle of the
filament is smaller than in the front (this would imply a shorter
time of excitation, or smaller $C_2^J$). But in any case, this
simple argument gives for the transition to two-humped filaments
the condition $w_s\approx w_f$, that is $\widetilde{\Da}\approx
C_2^J \epsilon^{-1}\approx 67$, to be compared with the actual
numerical value $\widetilde{\Da}\approx45$.

\begin{center}
\begin{figure}
\epsfig{file=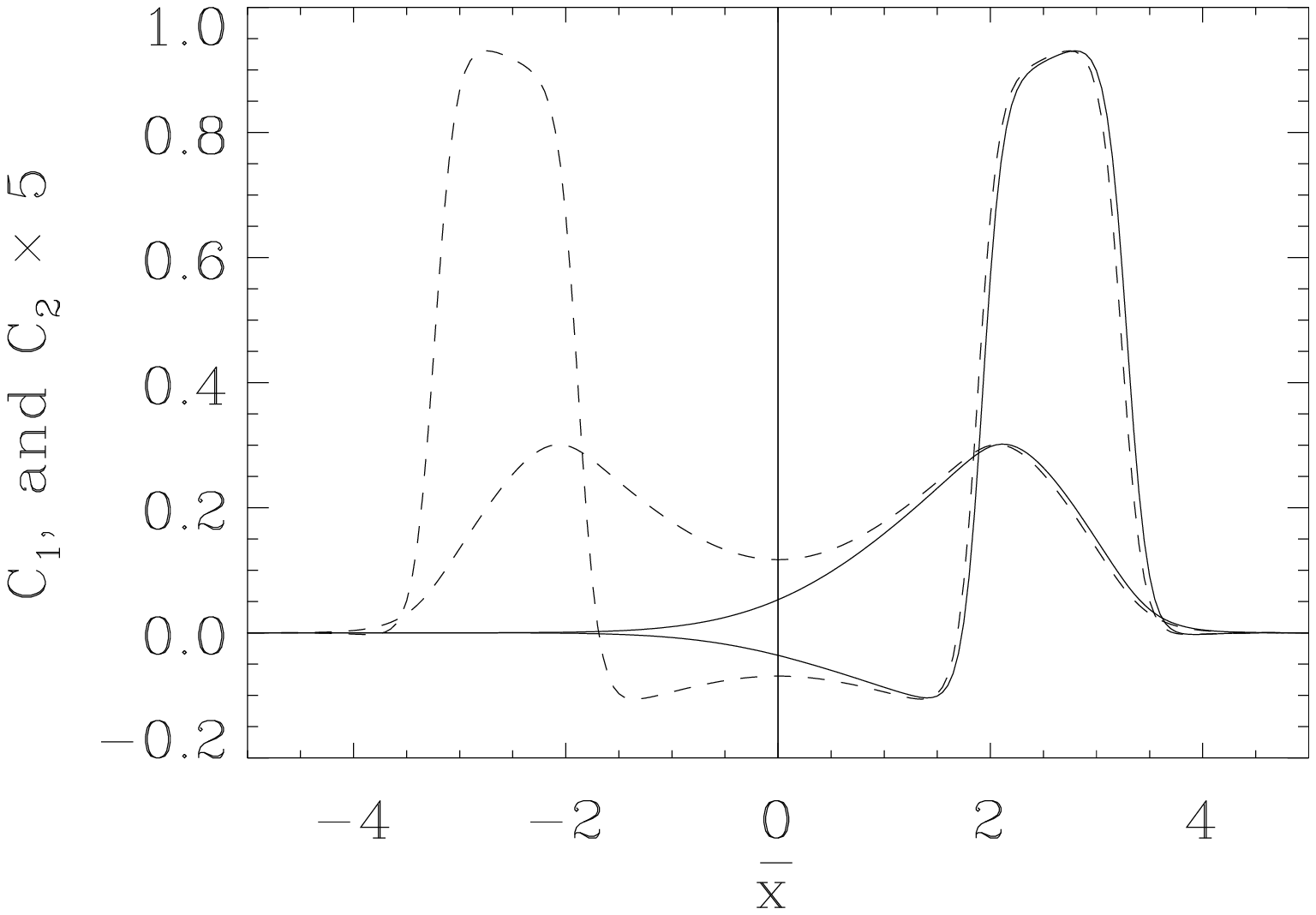,width=.50\textwidth}
\caption{Solid line is an asymmetric steady solution of Eq. (\ref{strain3}) for
$\widetilde{\Da}=200$. The higher curve is $C_1$ and the lower one
$C_2$, multiplied by a factor of 5 to be visible in this scale.
The dashed lines are symmetric solutions similar to the ones in
Fig. \ref{fig:2hfilament}, for $\widetilde{\Da}=200$ (upper curve,
$C_1$, lower curve, $C_2$ multiplied by 5). Numerically we find
that both solutions are simultaneously stable, each with its own
basin of attraction.}
\label{fig:asymmetric}
\end{figure}
\end{center}

We mention that, the transition from the unimodal steady solution
to the two-humped steady solution in the range 
$50 \lesssim \widetilde{\Da} \lesssim 70$ is associated with a
complex bifurcation scenario in which different stable solutions
coexist. 
For $\widetilde{\Da} \gtrsim 70$, additional asymmetric
steady stable solutions to Eq. (\ref{strain3}) appear. These are
similar to the pulse solutions obtained
without strain, but stopped by the flow. They have an excited
head, and a refractory tail. An example is shown in Fig.
\ref{fig:asymmetric}. The symmetric two-humped filaments found
before, which remain linearly stable, can be thought as bound
states of two asymmetric ones.

The simple model studied in this Section has allowed us to
qualitatively understand the individual filaments seen in the full
two-dimensional simulations, and in the baker model, to a great
detail. What is completely missed here is the interaction between
different filaments, or parts of the same filament. We have
studied filament collisions in the context of Eq. (\ref{strain3})
by initializing it with different combinations of displaced
filament solutions. The analogy with the collisions in the models
of the previous sections is far from complete, since here all the
filaments evolve in the same simple velocity field $-\lambda x$,
whereas in real multifilament situations, each filament has been
created around its own local strain. Nevertheless we have observed
that collision between symmetric one-humped filaments leaves at
long times a single centered one-humped filament, and collisions
between two-humped filaments annihilates half of the humps,
leading again to a single two-humped filament as the final state.
The asymmetric front-like filaments annihilate when colliding
front to front, and bind in a two-humped filament when colliding
tail to tail.

These observations help to understand the dynamic process of
filament merging that leads to the persistent patterns in the open
flows at intermediate $\Da$. With the consideration of periodic
boundary conditions, it is not difficult to understand also the
process ending excitation in the closed flow at not too large
$\Pe$ in terms of the annihilation of halves of two-humped
filaments.

What seems to escape from the picture is the process ending the
persistent pattern in the open flows at large $\Da$: because of
the openness of the flow, there are always filaments that do not
collide with others, but that receive the fresh reactants entering
the system. In this case, as in the situations of very large
$\Pe$, it seems that strain fluctuations are essential to
understand the process of deexcitation. We have run model
(\ref{strain}) with $\lambda$ randomly changing in time and found
that, for large enough fluctuations, nonvanishing correlation
time, and values of $\Da$ in the two-hump regime: the induced
width fluctuations eventually end with the decay of the filament,
in very much the same way as seen in the baker models simulations.
We speculate that the larger width fluctuations originated by
filament collisions in situations such as those in Fig.
\ref{fig:bakerOpen} (right panel) would amplify still more the
effect of unsteady strain and help to eliminate excitation. Strain
fluctuations in the baker model (periodic application of
contraction followed by periods without strain) are an artifact of
the discrete nature of the model. But in the two-dimensional
simulations of Sect. \ref{sec:numerics}, and in real flows, strain
fluctuations occur naturally and may be thus responsible for the
excitation decay in the open flow at large $\Da$. It is quite
natural that this decay process only appears after the filaments
develop the two-hump structure, since this implies the presence of
refractory material. Nevertheless, a quantitative description of
this process is still missing.

\section{Conclusions}
\label{sec:conclusions}

We have analyzed the behaviour of an excitable medium in the
presence of open or closed chaotic flows. In both cases, three
different regimes have been elucidated. The one at smaller $\Da$,
that is the dilution of the excitation at fast stirring, is
analogous to what is found in the case of bistable chemical
dynamics \cite{ZoltanFocus}.

The most interesting regimes are found at intermediate $\Da$: in
the closed flow case, a coherent excitation of the whole system
arises from the localized perturbation, whereas in the open flow
the excitation remains indefinitely in the system. This last
phenomenon was also found under bistable and in autocatalytic
dynamics \cite{ZoltanFocus}, as well as the excitation phase under
the closed flow, that is, the growth of an excited filament that
becomes space filling. What is distinct of the excitable dynamics
is that excitation under the closed flow is a transient, so that
the system finally recovers the rest state, at variance with the
bistable and autocatalytic behaviour. It is striking that this
recovery does not occur under the open flow in this intermediate
$\Da$ range.

Also a consequence of the recovery behaviour that characterizes the
excitable dynamics, and that distinguishes it from the otherwise rather
similar bistable dynamics, is the loss of coherence occurring at
large $\Da$. It manifests gradually under the closed flow, but as
a sudden disappearance of permanent excitation in the open case.

A great part of our work has been devoted to the development of
simplified models that help to understand the above regimes and
the transitions among them. Despite the strong approximations
performed, these simple models reproduce, at least qualitatively,
the full two-dimensional numerical results. The first simplified
model is based in the use of a baker map for the advection
dynamics. It highlights the processes of stretching and folding as
the basic flow mechanisms leading to the aforementioned chemical
regimes. The baker model dynamics also suggest that transitions
are linked to the properties of individual filaments. Thus, an
even simpler model is considered, where the stationary transverse
profile of a filament is the main quantity under study. Most of
the numerical observations can be understood within this
framework, although some phenomena, specially those for which
filament interactions seem relevant, would need of more detailed
modeling.

Some geophysical observations have been already interpreted within
the present framework\cite{GRL}. It would be of great interest to
perform experiments of chemical dynamics under well-controlled
stirring to observe the different scenarios predicted here.

\acknowledgements
C.L. acknowledges financial support from the Spanish MECD. E.H-G
acknowledges support from MCyT (Spain) projects BFM2000-1108
(CONOCE) and REN2001-0802-C02-01/MAR (IMAGEN).

\end{multicols}
\end{document}